\documentclass[twocolumn,aps,prl,showpacs,amsmath,amssymb,nobalancelastpage]{revtex4}
\usepackage{empheq}
\usepackage[dvips]{graphicx}
\usepackage{bm,eucal}
\usepackage{delarray,fancybox}
\usepackage{mathdots}
\usepackage[breaklinks,linktocpage]{hyperref}
\begin{document}
%
%
\newcommand{\nn}{\nonumber\\}
\newcommand{\rbr}[1]{\left(#1\right)}
\newcommand{\sbr}[1]{\left[#1\right]}
\newcommand{\cbr}[1]{\left\{#1\right\}}
\newcommand{\bra}[1]{\langle #1 |}                            
\newcommand{\ket}[1]{|#1 \rangle}                             
\newcommand{\av}[1]{\prec #1 \succ}
\newcommand{\dd}[1]{\eta_{#1}}
\newcommand{\eval}[2]{\left .#1\right|_{#2}}
\newcommand{\eq}[1]{(\ref{#1})}
\newcommand{\matgrap}[3]{\parbox{#2}{\includegraphics[width=#1]{#3}}} 
\newcommand{\ang}[1]{\frac{\Omega_{#1}}{(2\,\pi)^{#1}}}
\newcommand{\fou}[1]{\check{#1}}                             
\newcommand{\fvol}[1]{ \frac{d^{d} #1}{(2\,\pi)^{d}}}        
\newcommand{\dr}{\partial}                                   
\newcommand{\der}[2]{\frac{d #1}{d #2}}                      
\newcommand{\pder}[2]{\frac{\partial #1}{\partial #2}}       
\newcommand{\fde}[2]{\frac{\delta #1}{\delta #2}}            
\newcommand{\fset}[2]{\chi_{#1}\left(#2\right)}              
\newcommand{\dirac}[2]{\delta^{(#1)}(#2)}                    
\newcommand{\flu}[1]{\delta #1}                              
\newcommand{\vv}[1]{\boldsymbol{#1}}                         
\newcommand{\uv}[1]{\hat{\boldsymbol{#1}}}                   
\newcommand{\m}{\bar{m}}
\newcommand{\z}{\bar{z}}
\newcommand{\zs}{\bar{z}_{J,+}}
\newcommand{\zl}{\bar{z}_{J,-}}
\newcommand{\zi}{\bar{z}_{J,i}}
\newcommand{\is}{I_{+}}
\newcommand{\il}{I_{-}}
\newcommand{\ii}{I_{i}}
\newcommand{\dn}{d_{n}}
\newcommand{\ivec}{\mathcal{T}}
\newcommand{\GG}[1]{G_{#1}}
\newcommand{\eps}{\varepsilon}
\newcommand{\eddy}{\varkappa}
\newcommand{\hh}[1]{\mathcal{Y}_{#1}}
\newcommand{\hp}[1]{\mathcal{H}_{#1}}
\newcommand{\hf}[1]{\mathcal{K}_{#1}}
\title{On scaling and statistical geometry in passive scalar turbulence}
\author{Andrea Mazzino}
\affiliation{Department of  Physics, University of Genova, 
INFN and CNISM, via Dodecaneso 33, 16146 Genova, Italy.}

\author{Paolo Muratore-Ginanneschi}
\affiliation{Department of Mathematics and Statistics, 
University of Helsinki PL 68, 00014 Helsinki, Finland.}
%
\begin{abstract}
We show that the statistics of a turbulent passive scalar at scales larger than 
the pumping may exhibit multiscaling due to a weaker mechanism than the presence 
of statistical conservation laws. 
We develop a general formalism to give explicit predictions for the large scale 
scaling exponents in the case of the Kraichnan model 
and discuss their geometric origin at small and large scale. 
\end{abstract}
\maketitle


Turbulent transport poses challenges for fundamental research with important implications
for many environmental (e.g. 
impact of natural and anthropogenic pollutants on climate) and industrial 
(e.g. design of effective mixers of chemical products) applications. 
During the last fifteen years, the field has seen major developments \cite{FaSr06}.
The study of an analytical tractable model,  the Kraichnan model of passive advection 
\cite{Kr68,Kr94}, permitted for the \emph{first time} \cite{GaKu95,ChFaKoLe95} 
to prove that the statistics  of a turbulent passive field (e.g. the temperature) is 
\emph{intrinsically} not self-similar in the inertial range 
(fine scales of fluid motion not affected by thermal dissipation). 
More importantly, drawing on concepts and methods from stochastic analysis 
\cite{GaZe97,BeGaKu98} pointed out a general mechanism accounting for the experimentally 
and numerically observed multiscaling (see e.g. \cite{Wa00,ICTR08}) of inertial 
range statistical indicators. 
Accordingly, the statistics of equal time correlation functions is dominated 
by global statistical invariants of the Lagrangian dynamics \cite{BeGaKu98,FaGaVe01}. 
Although this picture can be established in a mathematically 
controlled way only for the Kraichnan model, numerical investigations of passive 
scalar advected by the Navier--Stokes equations \cite{CeVe01} together with 
experiments \cite{MyPuShSiWa98,Wa00} 
give strong evidences of the generality of the mechanism. 
In the unfolding of these developments, thoroughly summarized in \cite{FaGaVe01}, much 
attention has been devoted to the turbulent inertial range. However, in many physical contexts
(e.g. the study of the large scale structures in cosmology \cite{LiddleLyth}) it is important to 
understand the defining properties of statistical indicators of fluid tracers at scales larger 
than the typical energy source. 
As the energy of tracers transported by an incompressible velocity field 
is expected to ``cascade'' towards finer-scale, one might be tempted to infer from the absence  
of a ``constant-flux'' solution of the type predicted by Komogorov's 1941 theory \cite{Frisch} 
the onset of a thermodynamical equilibrium with Gaussian statistics and equipartition 
of scalar variance.
However it was recently shown analytically \cite{FaFo05} and numerically \cite{CeSe05,CeSe06} 
that the presence of an equipartition-like scalar power-spectrum may well co-exist 
with higher order correlation functions exhibiting breakdown of self-similarity and 
multiscaling.  Underlying these results is the existence, predicted in \cite{BeGaKu98} 
for the Kraichnan model, of an asymptotic 
zero-mode expansion of correlation functions \emph{also} at scales larger than the pumping. 
Here, we device a formalism
to calculate (perturbatively) 
the scaling dimensions of the large scale zero modes. We show that
large scale zero modes are not \emph{global} statistical conservation laws of the 
Lagrangian dynamics. 
They share however with inertial zero modes a geometrical origin indicated by their being 
in first approximation specified by eigenvalues of quadratic Casimir's of classical groups. 
Finally we provide numerical evidence of large scale zero mode dominance and discuss the 
relevance of these results for advection by Navier--Stokes.
The passive advection of a scalar quantity by a Newtonian incompressible 
fluid is governed by the equation
\begin{eqnarray}
\dr_{t}\theta+\boldsymbol{v} \cdot\boldsymbol{\dr}\, \theta-\frac{\kappa}{2}
\dr^{2}\theta=f
\label{Stratonovich}
\end{eqnarray}
where $\vv{v}$ is a vector field solving the Navier--Stokes equation and $f$ a stochastic 
large scale stirring. Following  Kraichnan \cite{Kr68,Kr94} we model 
turbulent fluctuations of $\vv{v} $ by a Gaussian statistics with zero average 
and 
\begin{eqnarray}
\av{v^{\alpha}(\vv{x},t)v^{\beta}(\vv{y},s)}=
\delta(t-s)\,D_{(\xi)}^{\alpha\,\beta}(\vv{x}-\vv{y},m)
\label{velocity}
\end{eqnarray}
where the spatial part of the velocity correlation is scale invariant up to an 
inverse integral scale $m^{-1}$. Such behaviour is encoded in 
the Mellin representation \cite{KuMG06}
\begin{eqnarray}
\lefteqn{\tilde{D}_{(\xi)}^{\alpha\,\beta}(\vv{x};m,z):=\int_{0}^{\infty}\frac{dw}{w}\,
\frac{D_{(\xi)}^{\alpha\,\beta}(w\,\vv{x};m)}{w^{z}}}
\nonumber \\
&&=-\frac{D_{0}\,\xi\,m^{z-\xi}\,\bar{C}(z,\xi) }{z-\xi}
\int \fvol{q}  \frac{e^{\imath \vv{q}\cdot \vv{x}}}{q^{d+z}}\,
\Pi^{\alpha\,\beta}(\uv{q})
\label{velocity_Mellin}
\end{eqnarray}
where $\Pi^{\alpha\,\beta}$ denotes the Fourier space transversal projector.
If $D_{(\xi)}^{\alpha\,\beta}$ decays faster than power-law for $m x\gg 1$ as we suppose here, 
$\bar{C}(z,\xi)$ is a meromorphic function analytic 
for $\Re z \in (-\infty,0)$ and analytic non-vanishing for $\xi\,\in\,[0,2)$.
The residues of the simple poles for $\Re z=0,\xi$  yield the inertial range asymptotics 
\cite{KuMG06}. 
For the statistics of the forcing field $f$ we hypothesise
time decorrelation (to preserve Galilean invariance), parity and translational invariance and 
correlation functions with support peaked around an integral scale $\m^{-1}\,\ll\,m^{-1}$. 
Mathematically, \eq{Stratonovich} is a stochastic partial 
differential in Stratonovich sense \cite{Ok} in order to preserve the hydrodynamic 
interpretation. A straightforward application of Ito lemma  (see e.g. \cite{FaGaVe01,KuMG06})
yields the Hopf equations satisfied by the scalar correlation function $\mathcal{C}_{n}$ of 
$n$-fields:
\begin{eqnarray}
\cbr{\dr_{t}-\frac{1}{2}\sum_{i\neq j}^{n}D_{(\xi)}^{\alpha\,\beta}(\vv{x}_{ij};m)
\dr_{x_{i}^{\alpha}}\dr_{x_{j}^{\beta}}-\frac{\eddy_{\kappa,m}^{(\xi)}}{2}\Delta_{n}}
\mathcal{C}_{n}=\mathfrak{F}_{n}
\label{Hopf}
\end{eqnarray}
with $\Delta_{n}$ the Laplacian in $\mathbb{R}^{n\,d}$,  $\vv{x}_{ij}:=\vv{x}_{i}-\vv{x}_{j}$, 
Einstein convention on contracted indices and $\mathfrak{F}_{n}$ an effective forcing depending 
at most on $\mathcal{C}_{n-2}$. The eddy diffusivity 
$\eddy_{\kappa,m}^{(\xi)}:=\kappa+{D_{(\xi)}}^{\alpha}_{\,\,\alpha}(0;m)/d$
has a finite inviscid limit $\eddy_{0,m}^{(\xi)}$ \emph{for all} $\xi \,\in\,[0,2]$. 
Translational invariance reduces the left hand side of \eq{Hopf} to 
$(\dr_{t}-M_{n}^{(\xi)})\mathcal{C}_{n}$ with $M_{n}^{(\xi)}$  a degenerate 
elliptic operator (for vanishing $\kappa$ and generic $\xi$)  
in $\dn:=(n-1)\,d$ spatial dimensions \cite{BeGaKu98}. 
The nullspace of $M_{n}^{(\xi)}$ can be thought as consisting of local martingales 
of an effective purely multiplicative stochastic process for each value of $n$. 
The relevance of these quantities for the unique solution \cite{Ha03} in 
$\mathbb{L}^{2}(\mathbb{R}^{\dn})$ of \eq{Hopf} is discussed in details in 
\cite{BeGaKu98, FaGaVe01}. 
The limit $\xi\downarrow 0$ illustrates the situation. In such a limit 
\cite{GaKu95} $D_{(0)}^{\alpha\,\beta}$ vanishes for
every finite point separation whilst still contributing to a scale independent 
inviscid eddy diffusivity $\eddy=\eddy_{0,m}^{(0)}$.  
Parametrising $\mathbb{R}^{\dn}$ with Jacobi variables (see e.g. \cite{FaGrBoHe99})
$\vv{R}=(\vv{r}_1,\dots,\vv{r}_{n-1})$, $\vv{W}=(\vv{w}_1,\dots,\vv{w}_{n-1})$, 
the reduction of the free 
Green function to the translational invariant sector 
reads \cite{Fa83}
\begin{eqnarray}
{M_{n}^{(0)}}^{-1}(\vv{R}-\vv{W})=
\sum_{J=0}^{\infty}\sum_{\vv{L}}
\frac{2\,\hf{J\vv{L}}(\vv{R})\,\hp{J\vv{L}}^{\dagger}(\vv{W})}{\eddy\,(\dn+2\,J-2)}
\label{Green_free}
\end{eqnarray}
for $R:=||\vv{R}||\geq W:=||\vv{W}||$. The $\hp{J\vv{L}}$'s are harmonic polynomials providing a 
complete orthonormal basis of $SO(\dn)$ through the relation
 $\hp{J\vv{L}}(\vv{R})=R^{J}\hh{J\vv{L}}(\uv{R})$  (here $\vv{R}:=R\,\uv{R}$) with 
hyperspherical harmonics labeled by $\dn-1$ integers $(J,\vv{L})$ (see e.g. \cite{FaGrBoHe99}). 
The $\hf{J\vv{L}}$'s are decaying harmonic functions in a one-to-one correspondence
with the $\hp{J\vv{L}}$'s specified by the so-called Kelvin transform \cite{AxBoRa01}:
\begin{eqnarray}
\hf{J\vv{L}}(\vv{R})=R^{2-\dn}\hp{J\vv{L}}\rbr{\vv{R}/R^{2}}
\end{eqnarray}
The $SO(\dn)$ decomposition of the Mellin transform of $\mathcal{F}_{n}$ 
\begin{eqnarray}
\tilde{\mathcal{F}}_{n}(\vv{R},\z)=\m^{-\dd{\mathcal{F}}}
\sum_{J\vv{L}}  (\m \,R)^{\z} \hh{J\vv{L}}(\uv{R}) F_{J\vv{L}}(\z)
\label{force_Mellin}
\end{eqnarray}
for $\dd{\mathcal{F}}$ the canonical dimension of $\mathcal{F}_{n}$ allows us to couch 
the steady state solution of \eq{Hopf} for vanishing $\xi$ as 
\begin{eqnarray}
\tilde{\mathcal{C}}_{n}^{(0)}(\vv{R},\z)=
\sum_{J\vv{L}}\frac{ 2\,\m^{-\dd{F}}R^{2}(\m \,R)^{\z}\,F_{J\vv{L}}(\z)\,\hh{J\vv{L}}(\uv{R})}
{\eddy\,(\dn+J+\z)\,(J-2-\z)}
\label{free_sol}
\end{eqnarray}
\eq{force_Mellin}, \eq{free_sol} can be thought as functionals of identical Lagrangian particles 
in the unique steady state. Thus there and in the following, for each $J\in \mathbb{N}$ 
the sum over $\vv{L}$ is restricted to fully symmetric states.
To each hyperangular sector is associted a  strip of analyticity, determined by the convergence 
of the Mellin integral, of size $-\dn-J\,<\,\Re \z \,<\,J-2$. The simple poles marking the 
boundary of the strip determine the non-canonical scaling dimensions of the large $\hf{J\vv{L}}$ 
and small scale $\hp{J\vv{L}}$ zero-modes. 
Thus, the expansion \eq{free_sol} evinces the geometrical  origin, $SO(\dn)$-anisotropy, 
of non-dimensional scaling.
Both classes of zero modes are local martingales as they belong to the nullspace of 
$\Delta_{n-1}$. 
However \emph{only} the $\hp{J\vv{L}}$ are \emph{strict} martingales i.e. are preserved by the 
propagator $P_{t}:=\exp(t\,\Delta_{n-1})$ of the diffusion: $\hp{J\vv{L}}=P_{t}\star \hp{J\vv{L}}$.
A direct calculation shows that projecting first $P_{t}$ onto its $(J\,,\vv{L})$-component 
renders the convolution $P_{t}\star \hf{J\vv{L}}$ integrable 
but restricts the region where the martingale property 
is satisfied to a domain $R^{2}\gg \eddy\,t$ 
monotonically decreasing in time. The $\hf{J\vv{L}}$ are therefore 
\emph{strictly local martingales} \cite{ElLiYo99}. 
The perturbative construction below in the text  suggests that large scale zero modes 
\emph{are not} expected in general to be statistical conservation laws of the dynamics.  
At small but finite $\xi$  the $SO(\dn)$-symmetry is broken to $\sigma_{n}\times SO(d)$ with 
$\sigma_n$ the permutation group of $n$ particles. 
As first shown in \cite{GaKu95} solutions of \eq{Hopf} can be constructed in a 
systematic perturbation theory in $\xi$. Combining \eq{Green_free} with \eq{force_Mellin} 
yields  for the $J\vv{L}$ component of  
$\mathcal{C}_{n}=\mathcal{C}_{n}^{(0)}+\xi\,\mathcal{C}_{n}^{(1)}+O(\xi^{2})$ in the steady state
\begin{eqnarray}
\lefteqn{\mathcal{C}_{n,J\,\vv{L}}^{(1)}(R,z,\z)=
-\frac{\mathcal{C}_{n,J\,\vv{L}}^{(0)}(R,\z)\,\ln m}{z}}
\nn&&
-\,\frac{2^{\frac{z}{2}}\,n\,(n-1)\,R^{2}(m\,R)^{z}\,(\m\, R)^{\z}\,C(z)}
{z^2\,\rbr{\dn+J+z+\z}\,\rbr{J-2-z-\z}}\times
\nn&&
\sum_{a=1}^{2}\int d\Omega_{\dn}\eval{\hh{J\vv{L}}^{\dagger}(\uv{W})\,\mathsf{J}_{aa}\,
\mathcal{D}_{a}\mathcal{C}_{n}^{(0)}(\vv{W},\z)}{\substack{W=1\\ \m=1}}
\label{sol_first_order}
\end{eqnarray}
with $\mathcal{D}_{a}:=w_{1}^{z}\cbr{\delta^{\alpha\beta}-\frac{z}{d-1+z} 
\frac{w_{1}^{\alpha}w_{1}^{\beta}}{w_{1}^{2}}}\dr_{w_{a}^{\alpha}}\dr_{w_{a}^{\beta}}$
and $C(z)$ such that $C(0)=1$. 
In deriving \eq{sol_first_order} we adopted an orthonormal set of Jacobi variables 
such that $\vv{r}_{1}:=\vv{x}_{12}$ and $\vv{r}_{2}:=\frac{(n-2)(\vv{x}_{1}+\vv{x}_{2})
-2\,\sum_{j=3}^{n}\vv{x}_{j}}{\sqrt{2\,(n-2)\,n}}$. 
In such a case the Jacobian of the change of variables give only two non-vanishing contributions 
$(\mathsf{J}_{11},\mathsf{J}_{22})$ equal to $(\frac{1}{2},\frac{n-2}{2\,n})$. 
The order of evaluation of the residues 
in the Mellin variables $z,\z$ determines the order of the limits of vanishing $m$ and $\m$. 
The condition
$m \ll \m$ is enforced evaluating first  the residue for $z$ equal zero.  
Corrections to scaling are then 
associated to \emph{double poles} in $\z$ occurring only for $\zs=J-2$ (inertial range) and  
$\zl=-\dn-J$ (large scales). 
Thus it is sufficient to diagonalise \eq{sol_first_order} in the
 $SO(\dn)$-representation specified by $J$. 
Universal terms in the two asymptotics, labeled by $i=\cbr{+,-}$, are encoded into 
finite dimensional matrices $\ii$ depending upon the asymptotics and the 
$SO(\dn)$-representation:
\begin{eqnarray}
&&\lefteqn{\hspace{-0.8cm}\mathcal{C}_{n;J\vv{L}}^{(1)}(\vv{R},i)\to
\frac{2\,\m^{-\dd{\mathcal{F}}} R^{2+\zi}}{\eddy\,(\dn+2\,J-2)}\left\{
F_{J\vv{L}}(\zi)\ln\frac{\m}{\sqrt{2}}\right. }
\nn&&\hspace{-0.7cm}
\left. -(-1)^{i}\ln(\m\, R)\sum_{\vv{L}^{\prime}}\bra{J,\vv{L}}\ii \ket{J,\vv{L}^{\prime}}
F_{J\vv{L}^{\prime}}(\zi)\right\} +\dots
\label{sol_first_order_red}
\end{eqnarray}
The ``$\dots$'' stand for non-logarithmic corrections. Scaling exponents are determined 
by the eigenvalues
$\zeta_{\zi}^{(1)}$ of $I_{i}$ according to $\zeta_{\zi}=2+\zi+\xi\,\zeta_{\zi}^{(1)}+O(\xi^{2})$.
It is expedient to choose a representation of hyperspherical harmonics adapted to 
the group-subgroup chain 
adapted to $SO(\dn)\supset SO(d)^{n-1}$ (see e.g. \cite{Fa83,FaGrBoHe99}). 
If we focus on the $SO(d)$-isotropic sector of $\mathcal{C}_{4}$ as in \cite{GaKu95} for
permutation invariant states the representation is two-dimensional and all calculations 
can be performed explicitly \cite{footnote}.
The inertial range asymptotics recovers the results
\begin{eqnarray}
\zeta_{4,+}^{{(1)}}([4,0])=-\frac{2\,(d+4)}{d+2}\,,
\hspace{0.3cm}
\zeta_{4,+}^{(1)}([4,2])=-\frac{2\,(d-2)}{d-1}
\label{small_scale}
\end{eqnarray}
respectively corresponding to the irreducible and reducible zero modes \cite{GaKu95}.
The  large scale asymptotics yields
\begin{eqnarray}
\zeta_{4,-}^{{(1)}}([4,0])=\frac{d+6}{d+2}\,,
\hspace{0.3cm}
\zeta_{4,-}^{(1)}([4,2])=\frac{d-3}{d-1}
\label{large_scale}
\end{eqnarray}
In order to interpret the results and justify the notation, 
we rewrite the scalar products on the $\dn$-hypersphere in \eq{sol_first_order} in terms of the 
Gaussian measure of $\mathbb{R}^{\dn} $ so that
for any $\eps>0$
\begin{eqnarray}
\lefteqn{\bra{J,\vv{L}}\il \ket{J,\vv{L}^{\prime}}=\sum_{a=1}^{2}\frac{2\,n\,(n-1)\,
\mathsf{J}_{aa}}{\dn+2\,J-2}\eval{\frac{d}{dz}}{\substack{z=0 \\ \z=-\dn-J}}\times}
\nn&&\hspace{-0.4cm}
\int d^{\dn}W\,\frac{e^{-\frac{W^{2}}{2\,R_{o}^2}}\,W^{\eps}\,
\hp{J\vv{L}}^{\dagger}(\vv{W})\mathcal{D}_{a}\,W^{2+\z}
\hh{J\vv{L}^{\prime}}(\uv{W})}{(2\, R_{o}^{2})^{\frac{z+\z+J+\eps}{2}}\,
\Gamma\rbr{\frac{z+\z+J+\eps}{2}}}
\label{lift}
\end{eqnarray}
so that we can integrate by parts in Cartesian coordinates. By incompressibility of 
\eq{velocity} the operation reduces to letting  $\mathcal{D}_{a}$ act to its left in \eq{lift}. 
Projecting back to the $SO(\dn)$ scalar product and taking the limit of vanishing $\eps$ 
yield the relation
$\bra{J,\vv{L}}\il \ket{J,\vv{L}^{\prime}}=\bra{J,\vv{L}^{\prime}}\is-1 \ket{J,\vv{L}}$
implying 
$\zeta_{\zl}^{(1)}=-\zeta_{\zs}^{(1)}-1$
satisfied by \eq{small_scale},\eq{large_scale} so that
$\zeta_{\zl}+\zeta_{\zs}=2-\dn-\xi+O(\xi^{2})$ which is consistent with the 
non-perturbative analysis of \cite{BeGaKu98}. 
In the literature (see e.g. \cite{BeGaKu96,AdAnVa98})
the $\zeta_{\zs}$'s have been computed in general for irreducible zero modes 
\cite{GaKu95,FaGaVe01} as they are the only to contribute to structure functions.
Here we outline a different approach  based on the martingale property of the $\hp{J\vv{L}}$'s
and conceptually ``dual'' to the Wilsonian renormalisation of composite operators of \cite{KuMG06}.
Instead of studying operators of the renormalised theory with larger infra-red cut-off we study 
martingales of the original theory in the limit of infinite integral scale. 
To this goal we introduce the infra-red regularised harmonic polynomials 
$\hp{J\vv{L}}^{[L]}(\vv{R}):=\hp{J\vv{L}}(\vv{R})\exp\{-R^{2}/(2L^{2})\}$.
These are eigenstates of the isotropic harmonic oscillator in $\mathbb{R}^{\dn}$ 
and, consequently, 
eigenstates of the Fourier transform \cite{Ro05}. 
Using this property and the diagrammatic techniques 
of \cite{KuMG06} it is straightforward  to evaluate the convolutions
\begin{eqnarray}
\lim_{L\uparrow \infty}{M_{n}^{(0)}}^{-1}\star \frac{\hp{J\vv{L}}^{[L]}}{L^{2}}=
\frac{2\,\hp{J\vv{L}}}{\eddy\,(\dn+2\,J-2)}
\label{hp_free}
\end{eqnarray}
and for $J>0$
\begin{eqnarray}
\lefteqn{\lim_{L\uparrow \infty}{M_{n}^{(1)}}^{-1}(z)\star \frac{\hp{J\vv{L}}^{[L]}}{L^{2}}=
-\frac{2\,\hp{J\vv{L}}\,\ln m}{z\,\eddy\,(\dn+2\,J-2)}}
\nn&&
-\sum_{l \neq k}\sum_{a,b}^{n-1}\mathsf{J}_{a1}^{(lk)}\mathsf{J}_{b1}^{(lk)}
\frac{\dr_{r_{a;(lk)}^{\alpha}}\dr_{r_{b;(lk)^{\beta}}}\,\hp{J\vv{L}}(\vv{R}_{(lk)})}
{z\,\varkappa\,(\dn+2\,J-2)}
\nn&&
\times\frac{d\,\bar{C}(z,0)\,m^{z}}{(d-1)}\int \fvol{q}\frac{2^{2+\frac{z}{2}}\,
e^{\imath\,\vv{q}\cdot\vv{r}_{1;(lk)}}\,\Pi^{\alpha\,\beta}(\uv{q})}{\bar{C}(0,0)\,q^{d+z+2}}
\label{hp_correction}
\end{eqnarray}
$\mathsf{J}^{(lk)}$ is the Jacobian of orthonormal Jacobi coordinates adapted to
$\vv{r}_{1;(lk)}=\vv{x}_{lk}/\sqrt{2}$. The integral in \eq{hp_correction} yields the first 
term of the loop expansion to which the perturbative theory for the $\mathcal{C}_{n}$'s 
reduces if the limit $\m \downarrow 0$ is taken first. 
The integral may seem to require analyticity of $\bar{C}(z,0)$ in the strip 
$\Re z\in [-2,0)$. However the residue for $\Re z=-2$ is proportional to 
$\Delta_{n-1} \hp{J\vv{L}}$ and vanishes. The scaling
dimensions of the inertial range zero modes are determined by prefactor of 
the self-similarity breaking term $\ln m$. After some algebra we get into
\begin{eqnarray}
&&\hspace{-0.5cm}\mathrm{Res}_{z=0}\cbr{\lim_{L\uparrow \infty}{M_{n}^{(1)}}^{-1}(z)
\star \frac{\hp{J\vv{L}}^{[L]}}{L^{2}}}=
\frac{2\,\ln m}{\eddy\,(\dn+2\,J-2)}\times
\nn&&
\cbr{1+\frac{(d+1)\,\mathfrak{C}_{SO(d)}^{(2,n)}-d\,
\mathfrak{C}_{SU(d)}^{(2,n)}}{2\,(d-1)\,(d+2)}}\hp{J\vv{L}}+\dots
\label{pt_result}
\end{eqnarray}
with $\mathfrak{C}_{SU(d)}^{(2,n)}=\mathfrak{C}_{SU(n-1)}^{(2,n)}
+\frac{(d+1-n)}{d\,(n-1)}\mathfrak{E}(\mathfrak{E}+\dn)$, 
$\mathfrak{E}$ the generator of dilations 
and $\mathfrak{C}_{SO(d)}^{(2,n)}$, $\mathfrak{C}_{SU(n-1)}^{(2,n)}$ the, mutually 
commuting, quadratic Casimir invariants of $SO(d)$, $SU(n-1)$ acting on translation and
permutation invariant homogeneous polynomials of $n$-particle variables in $d $-dimensions. 
Although 
$[\Delta_{n},\mathfrak{C}_{SU(n-1)}^{(2,n)}]\neq 0$, any homogeneous polynomial $P$ 
of degree $J$ admits a unique expansion $P_{J}=\sum_{k=0}^{k_{\star}}R^{2\,k}H_{J-2k}$,
$k_{\star}=\mathrm{int}(J/2)$ for the $H_{J}$'s harmonic homogeneous polynomials of degree $J$ 
\cite{AxBoRa01}. Thus linear combinations of the  $\hp{J\vv{L}}$'s specify eigenstates of 
the Casimir invariants up to \emph{slow modes} of the free theory \cite{BeGaKu98}. 
By Gel'fand-Zetlin theory (see e.g.\cite{Lo70}) the 
eigenvalues are $\lambda_{SO(d)}(j)=j\,(j+d-2)$ and 
$\lambda_{SU(n-1)}(\vv{a})=\sum_{i=1}^{n-2} a_{i}\rbr{a_{i}-2\, i}+\frac{J[(n-1)n-J]}{n-1}$
for $j$, $\boldsymbol{a}=[a_{1},\dots,a_{n-2}]$ non-negative integers satisfying
$\sum_{i=1}^{n-2}a_{i}=J$ and $a_{i}\geq a_{j}$ for any $i\geq j $ 
so that:
\begin{eqnarray}
\lefteqn{\zeta_{\zs}^{{(1)}}(j,\vv{a})=\frac{(d+1)\,j\,(j+d-2)}{2\,(d-1)\,(d+2)}}
\nonumber\\
&&-\frac{d \sum_{i=1}^{n-2} a_{i}\rbr{a_{i}-2\, i}+J[d(d+1)-J]}
{2\,(d-1)\,(d+2)}
\label{ss_exponent}
\end{eqnarray}
Irreducible zero modes correspond to $\boldsymbol{a}=[n,0,\dots,0]$ ( $J=n$ and $n-3$ zeroes) 
whilst the four point reducible zero mode to $\boldsymbol{a}=[2,2]$.
For  $\mathcal{C}_{2}$ \cite{FaFo05,CeSe05,CeSe06}  
the value of the forcing spectrum at zero momentum determines whether the decay at scales 
larger than the pumping is power law or exponential, in the latter case paving the way for 
anisotropic scaling dominance. 
Fig.~\eq{fig:1} illustrates realizability of large scale anomalous scaling 
for $\mathcal{C}_{4}$ and non Gaussian forcing.
\begin{figure} [h!]
\centerline{
\includegraphics[scale=0.6,draft=false,angle=0]{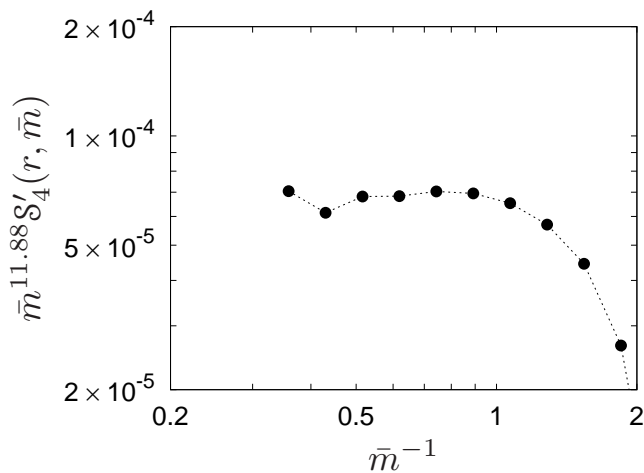}}
\caption{Numerical large scale behaviour of 
$\mathcal{S}_{4}^{\prime}(r,\m):=\mathcal{S}_{4}(r,\m)-\mathcal{C}_{4}(0,\m)$  
($4$-point structure function minus $4$-point correlation at coinciding points)
versus the integral scale $\m^{-1}$ balanced by the theoretical zero mode prediction 
$\m^{\zeta_{4,-}^{\prime}([4,0])}$ with $\zeta_{4,-}^{\prime}([4,0])=2-\xi-\zeta_{4,-}([4,0])$.
The plot is obtained by averaging over $N=10^{9}$ 
lagrangian paths using the algorithm of \cite{FrMaVe98} 
at $\xi=0.4$, $r=1$ and $d=3$. By \eq{large_scale} $\zeta_{4,-}^{\prime}([4,0])=11.88+O(\xi^{2})$. 
Forcing is non-Gaussian and proportional to the hyperspherical harmonic $\hh{4,\vv{L}^{\star}}$ 
specifying the zeroth order of the \emph{irreducible} inertial range zero mode 
(see \cite{GaKu95} for details). 
The observed behaviour significantly deviates from the scaling prediction coming from 
the exponent $2-d_{4}-\xi$ of the Green function.
}
\label{fig:1}
\end{figure}

These results give an analytical though perturbative validation of the 
general link between geometry and intermittency in passive scalar turbulence 
numerically established in \cite{CeVe01}. Furthermore, in the inertial range the above analysis 
carries over to a passive scalar advected by the Navier--Stokes equation in the 
thermal stirring regime forced by a Gaussian random field self-similar 
with H\"older exponent $\eps$. 
As shown in \cite{AdAnHoKi05}, at leading order in a loop expansion in $\eps$ the scalar 
is driven only by the Gaussian core of the velocity statistics described
by a Kraichnan model with $\xi \propto \varepsilon $.

This paper greatly benefited of many discussions
with A.~Kupiainen. We thank D.~Gasbarra 
for useful comments. 
This work was supported by the CoE {\em ``Analysis and Dynamics''} of the Academy 
of Finland.

\end{document}